\documentclass[prd,preprint,nofootinbib,superscriptaddress]{revtex4}

\usepackage{graphicx, cancel}
\usepackage{amsmath,amssymb,graphicx,psfrag}

\begin{document}

\newcommand{\gsim}{ \mathop{}_{\textstyle \sim}^{\textstyle >} }
\newcommand{\lsim}{ \mathop{}_{\textstyle \sim}^{\textstyle <} }

%%%%%%%%%%%%%%%%%%%%%%%%%%%%%%%%%

\renewcommand{\thefootnote}{\fnsymbol{footnote}}

\preprint{CERN-PH-TH/2009-024}
\preprint{DESY 09-027}

\title{R-parity Violating Right-Handed Neutrino \\
in Gravitino Dark Matter Scenario}

\renewcommand{\thefootnote}{\alph{footnote}}

\author{Motoi Endo}
\affiliation{
Theory Division, PH Department, CERN, CH-1211 Geneva 23, 
Switzerland
}
\author{Tetsuo Shindou}
\affiliation{
Deutsches Elektronen Synchrotron DESY, Notkestrasse 85,
22607 Hamburg, Germany
}

\begin{abstract}
\noindent
A decay of the gravitino dark matter is an attractive candidate to explain the 
current excesses of the PAMELA/ATIC cosmic-ray data. However, R-parity 
violations are required to be very tiny in low-energy scale. We suggest a 
R-parity violation in the right-handed neutrino sector. The violation is suppressed 
by a see-saw mechanism. Although a reheating temperature is constrained 
from above, the thermal leptogenesis is found to work successfully with a help 
of the R-parity violating right-handed neutrino. 
\end{abstract}

\maketitle

%%%%%%%%%%%%%%%%%%%%%%%%%%%%%%%%%
\section{Introduction}
%%%%%%%%%%%%%%%%%%%%%%%%%%%%%%%%%

Observations of comic rays have been greatly developed. After a series of 
cosmic-ray measurements, PAMELA recently published the first result of 
the positron fraction, which show an excess compared to the background 
at energies above 10GeV \cite{Adriani:2008zr}. Interestingly, the ATIC 
collaboration also reported an excess of the electron plus position flux in a 
range of $O(100)$GeV \cite{:2008zzr}. Although the experimental data and 
the background estimations still involve ambiguities, the excesses may be 
a sign of the dark matter (DM). In this letter, we focus on decaying DM 
scenarios. The DM is not always required to be stable as long as its life 
time is long enough for the DM to survive until today. Then, its decay products 
are a possible source of the energetic cosmic rays. 

It is non-trivial to realize such a long-lived particle. Actually, decay operators 
must be incredibly suppressed for the models to be viable. A well-motivated 
candidate of the DM is known to be the gravitino with a broken R-parity. 
This is because its decay is doubly suppressed by the Planck scale and 
R-parity violations \cite{Takayama:2000uz}. Even with the Planck suppression, 
the R-parity violation is limited to be very tiny and looks unnatural unless 
some mechanisms are taken into account. As a solution of the fine-tuning, 
we propose a R-parity violation introduced in the right-handed neutrino sector. 
Then, the violation is suppressed in the low-energy scale by a see-saw 
mechanism. Actually, since the violation appears in low-energy phenomena 
though the right-handed neutrino, its effect becomes suppressed by the 
right-handed neutrino mass scale. We will show that the PAMELA/ATIC 
excesses are explained for $M_N \sim 10^9$GeV without a parameter tuning. 

This type of the R-parity violation does not only explain the PAMELA/ATIC 
anomalies, but also opens a window for the successful thermal leptogenesis 
\cite{Fukugita:1986hr} especially in a large gravitino mass region. Indeed, the 
mechanism is an attractive and elegant solution to the mystery of the baryon 
asymmetry of the universe, though it usually requires a relatively high reheating temperature, $T_R \gsim 10^9$GeV (see e.g. \cite{Buchmuller:2004nz}). 
In such a high reheating temperature, the severest cosmological constraints 
are provided by the big-bang nucleosynthesis (BBN) and the overclosure of 
the universe. The former constraint is due to a prolonged decay of the 
next-to-lightest superparticle (NLSP) into the gravitino lightest superparticle 
(LSP). In contrast, the NLSP can decay into the Standard Model (SM) particles 
much more efficiently through the R-parity violating operator. Thus, the BBN 
bound becomes ameliorated \cite{Buchmuller:2007ui}. 

The overclosure constraint then puts a constraint on the reheating temperature. 
In particular, the thermal gravitino production is enhanced in a larger soft mass 
region. According to PAMELA/ATIC, the cosmic-ray excesses imply a heavy 
gravitino, namely large superparticle masses. We will see that the leptogenesis 
temperature is unlikely to be satisfied for $m_{3/2} \gsim 1$TeV unless the 
gluino mass parameter is relatively small at the GUT scale. Nevertheless, since 
the R-parity violation in this letter contributes to a CP-asymmetric right-handed 
neutrino decay \cite{Farzan:2005ez}, the thermal leptogenesis will be shown to 
work to explain the observed baryon asymmetry easily. Thus, we will study 
the R-parity violating scenario in the right-handed neutrino sector as an attractive 
framework of the decaying gravitino DM.

%%%%%%%%%%%%%%%%%%%%%%%%%%%%%%%%%
\section{R-parity violation in Right-Handed Neutrino Sector}
%%%%%%%%%%%%%%%%%%%%%%%%%%%%%%%%%

Let us consider the gravitino DM scenario with a decay due to R-parity violations. 
A spectrum of the cosmic rays from its decay products are determined by the 
type of the R-parity violating operator, the gravitino mass and its life time. In a 
case of the bilinear R-parity violation, the dominant decay channels are $\psi_{3/2} 
\to W\ell$, $Z\nu$, $\gamma\nu$ and $h\nu$ \cite{Ibarra:2007wg,Ishiwata:2008cu,Ibarra:2008qg}. 
The decay is parametrized by the sneutrino vacuum expectation values (VEV). 
With the operator, $W = \mu' LH$, it becomes 
\begin{equation}
  \langle \tilde \nu \rangle \;\simeq\; 
  - \frac{\mu^{\prime*} \mu v \cos\beta}{m_{\tilde\nu}^2},
\end{equation}
where $L$ and $H$ are the left-handed lepton and up-type Higgs, respectively. 
The parameters are the higgsino mass $\mu$, the Higgs VEV $v \simeq 174$GeV, 
the Higgs VEV ratio $\tan\beta = \langle H \rangle/\langle \bar H \rangle$, and the 
sneutrino soft mass $V_{\rm soft} = m_{\tilde \nu}^2 \tilde \nu^* \tilde \nu$. Then, 
the gravitino life time is estimated as 
\begin{equation}
  \tau_{3/2} \;\simeq\; 1 \times 10^{26}{\rm sec} 
  \left(\frac{\eta}{10^{-10}}\right)^{-2}
  \left(\frac{m_{3/2}}{1{\rm TeV}}\right)^{-3}
\end{equation}
for $m_{3/2} \gg 100$GeV, where $\eta$ is defined as $\eta \equiv 
|\langle \tilde\nu \rangle|/v$. According to the current PAMELA/ATIC data, 
the mass and life time of the gravitino is favored to be 
\begin{equation}
  m_{3/2} \;\gsim\; O(100-1000){\rm GeV},~~~~~~~~
  \tau_{3/2} \;\sim\; 10^{26}{\rm sec}.
  \label{eq:PAMELA-ATIC}
\end{equation}
which leads to the R-parity violation parameter as
\begin{equation}
  \eta \;\sim\; 10^{-10} \left(\frac{m_{3/2}}{1{\rm TeV}}\right)^{-3/2}.
\end{equation}
This $\eta$ satisfies with the other cosmological constraints \cite{Ishiwata:2008cu}. 
The gravitino survives until today as $\eta \ll 10^{-(5-6)}$ for $m_{3/2} = 
O(100)$GeV, and the parameter is small enough to avoid the wash-out of the 
lepton/baryon asymmetry since $\eta < 10^{-7}$ is satisfied. Furthermore, 
the BBN constraint is solved for $\eta \gsim 10^{-(11-12)}$ by virtue of the 
R-parity violation, and the R-parity violating contribution with $\eta < 10^{-7}$ 
is negligible for the neutrino mass. We assume that the neutrino mass is obtained 
by the see-saw mechanism in this letter. 

However, from the naturalness point of view, the violation is too tiny and must be 
fine-tuned. We propose a simple framework to realize this tiny R-parity violation. 
We introduce the following R-parity violating operator in the right-handed neutrino 
sector, 
\begin{equation}
  W \;=\; \lambda N H \bar{H}.
  \label{eq:RHNU-RPV}
\end{equation}
Accompanied by the right-handed neutrino mass and the Yukawa terms, 
$W = M_N NN/2 + Y_N NLH$, after decoupling the heavy right-handed neutrino, 
we obtain the effective R-parity violating operators in low-energy scale:
\begin{equation}
  W \;=\; 
  - \frac{\lambda Y_N (LH) (H\bar{H})}{M_N}.
  \label{eq:low-RHNU-RPV}
\end{equation}
It is noticed that the R-parity violating effects appear with $1/M_N$ because 
they contribute through the right-handed neutrino. Thus, the R-parity violating 
effects are naturally suppressed by the see-saw mechanism. 

Taking the Higgs VEVs, (\ref{eq:low-RHNU-RPV}) behaves as a bilinear R-parity 
violating operator. Actually, combined with the $\mu$ term and the sneutrino soft 
mass, the sneutrino acquires VEV as 
\begin{equation}
  \langle \tilde \nu \rangle \;\simeq\; 
  - \frac{\lambda Y_N \mu v^3 \sin^3\beta}{M_N m_{\tilde\nu}^2}.
  \label{eq:snuVEV}
\end{equation}
Then, $\eta$ is estimated to be
\begin{equation}
  \eta \;\simeq\; 
  0.6 \times 10^{-10} \cdot
  \lambda
  \left(\frac{M_N}{10^9{\rm GeV}}\right)^{-\frac{1}{2}}
  \left(\frac{m_{\tilde\nu}}{1{\rm TeV}}\right)^{-2}
  \left(\frac{\mu}{1{\rm TeV}}\right)
  \left(\frac{\bar{m}_\nu}{0.1{\rm eV}}\right)^{\frac{1}{2}}
  \sin^2\beta,
  \label{eq:eta-RHNU-RPV}
\end{equation}
where $\bar{m}_\nu$ is a typical scale of the light neutrino masses defined 
as $Y_N \equiv \sqrt{M_N \bar{m}_\nu}/v\sin\beta$. Here, we omitted flavor 
indices. As a result, we find that the gravitino life time $\tau_{3/2} \sim 
10^{26}$sec required from PAMELA/ATIC is realized for $\lambda \sim 1$ and 
$M_N \sim 10^9$GeV. We want to emphasize that there is no fine-tuning in the 
R-parity violating parameters. 

The gravitino decay is almost the same as that by the bilinear R-parity violation. 
In fact, it is dominantly induced by the sneutrino VEV (\ref{eq:snuVEV}), while 
the other decaying channels from (\ref{eq:low-RHNU-RPV}) are subdominant. 
Although $W = -\lambda^2 (H\bar{H})^2/2M_N$ is also derived from 
(\ref{eq:RHNU-RPV}), the resultant decay operators are negligible for the 
cosmic ray spectra. 

According to the current cosmic-ray data, the positron spectrum has a steep rise 
in high-energy region. This behavior prefers the gravitino DM decaying direct into 
the positron or anti-muon \cite{Ishiwata:2008cv,Ibarra:2008jk}. The decay product 
is determined by flavour structure of the sneutrino VEV. Denoting the flavour 
indexes explicitly in (\ref{eq:snuVEV}), $\langle \tilde \nu_j \rangle$ is proportional 
to $\lambda_i (Y_N)_{ij} /M_{Ni}$ with respect to the flavor structure, where $i$ is 
the index of the heavy neutrino and $j$ for the light one. Thus, it is not surprising 
to expect the positron and anti-muon to be produced in the decay. Consequently, 
the R-parity violating operator (\ref{eq:RHNU-RPV}) can naturally explain the current 
PAMELA/ATIC excesses with $m_{3/2} = O(100-1000)$GeV. 

Let us comment on radiative corrections with the R-parity violation 
(\ref{eq:RHNU-RPV}). They induce a linear term of the right-handed neutrino 
in the K\"ahler potential. This then leads to the right-handed sneutrino VEV in 
the framework of the supergravity. Although this can give a similar contribution 
to the effective R-parity violating operators (\ref{eq:low-RHNU-RPV}), the results 
in this letter do not change qualitatively. Thus, we will neglect it hereafter. 

We may also have the R-parity violating operators in the right-handed neutrino 
sector other than (\ref{eq:RHNU-RPV}) such as the superpotential terms, $N$ 
and $N^3$. However, their couplings are required to be suppressed since they 
can induce too large R-parity violations through the right-handed sneutrino VEV 
within the supergravity. Thus, we consider that the operators are forbidden by 
(discrete) symmetries. 

In the above, we focused on the SUSY invariant operators for the R-parity 
violations. Apart from them, we may have the violations in the SUSY breaking 
term. It can be noticed that the analysis is similar to the above, and the result 
is almost the same.

%%%%%%%%%%%%%%%%%%%%%%%%%%%%%%%%%
\section{Upper Bound on Reheating Temperature}
%%%%%%%%%%%%%%%%%%%%%%%%%%%%%%%%%

In the decaying gravitino DM scenario, the cosmological constraint comes 
from the overclosure of the universe. Actually, the gravitino is thermally 
produced in the hot plasma. At the leading order, the relic abundance is 
evaluated as
\begin{equation}
  \Omega_{3/2}h^2
  \;\simeq\;
  \sum_{i=1}^3 
  \omega_i g_i^2(T_R)
  \ln\frac{k_i}{g_i(T_R)}
  \left(1+\frac{M_i^2(T_R)}{3m_{3/2}^2}\right)
  \left(\frac{m_{3/2}}{100\text{GeV}}\right)
  \left(\frac{T_R}{10^{10}\text{GeV}}\right),
  \label{eq:gravitino-abundance}
\end{equation}
where the definition of the parameters are found in \cite{Pradler:2006qh}. 
In the numerical analysis, we include the electroweak contributions, which 
can be sizable especially when the gluino mass parameter is rather small 
compared to the Bino and Wino ones at the GUT (namely $T_R$) scale. 
It should be mentioned that the gravitino production rate includes an $O(1)$ 
uncertainty from unknown higher order contributions and nonperturbative 
effects \cite{Bolz:2000fu}. In addition, resummation of thermal masses 
potentially increases the rate by about a factor of two  
\footnote{
  Apart from the thermal production, we also have nonthermal 
  contributions to the gravitino abundance, particularly from inflaton 
  decay \cite{Endo:2007sz}. Since they are model dependent, we 
  neglect them for simplicity. 
} \cite{Rychkov:2007uq}. Since the massive gravitino behaves as a cold dark 
matter, its abundance is constrained from the measurements. According to the 
WMAP 5-year data, the abundance is required to satisfy \cite{Hinshaw:2008kr}, 
\begin{equation}\label{dark}
  \Omega_{3/2}h^2 
  \;\leq\; 
  \Omega_{\rm DM}h^2 
  \;\simeq\; 
  0.1223,
\end{equation}
at the 2$\sigma$ level. Thus, provided the gaugino and gravitino masses, 
we obtain an upper bound on the reheating temperature. 

It is interesting to study the overclosure bound in terms of the superparticle 
masses at the weak scale, which are expected to be measured by LHC. 
From (\ref{eq:gravitino-abundance}) we notice that the gravitino production 
is evaluated with the parameters at the reheating temperature scale. Then, 
they are correlated with those at the weak scale by solving the renormalization 
group equations. Let us mention that the 1-loop result includes large uncertainties 
especially in the colored sector: 2-loop contributions can give an $O(10)$\% 
correction, and the renormalization scale at the 1-loop level potentially contains 
additional $O(10)$\% uncertinty, which is reduced by taking the higher order 
corrections into account. Noting that (\ref{eq:gravitino-abundance}) includes 
the gluino mass squared, the thermal gravitino abundance can change drastically. Actually, $\Omega_{3/2}h^2$ is found to be almost as twice as the previous 
1-loop analyses for $M_i(T_R) \gg m_{3/2}$ with the gluino mass fixed at the 
weak scale \cite{Bolz:2000fu} (see also the discussion in \cite{Buchmuller:2008vw}). 
In the following numerical analysis, we evolve the renormalization group running 
at the 2-loop accuracy and include the 1-loop threshold corrections by means of 
{\tt SOFTSUSY} 2.0.18 \cite{Allanach:2001kg}. 

\begin{figure}
  \includegraphics[scale=0.7]{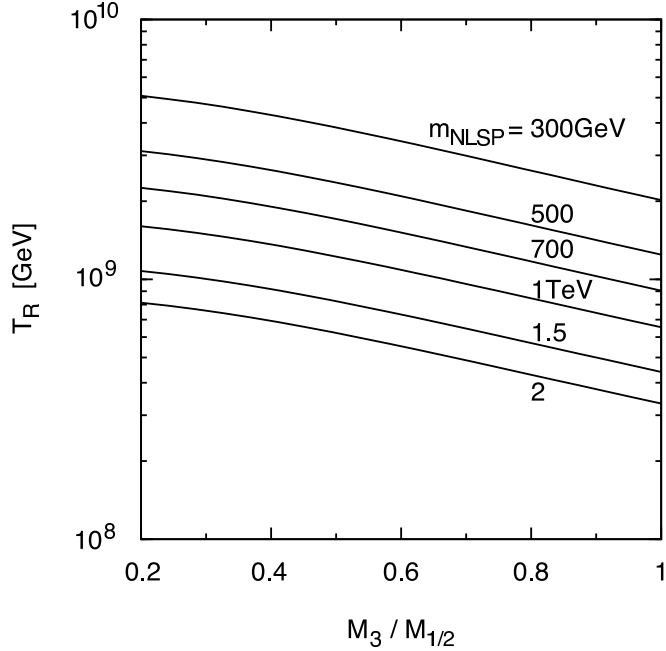}
  \caption{
    The maximal reheating temperatures allowed by the overclosure bound 
    are drawn for fixed NLSP masses. The gravitino mass is set to be equal 
    to the NLSP mass. Note that these masses are the physical one, while 
    $M_3$ is given at the GUT scale, and the soft scalar masses are chosen 
    to be $m_0 = M_{1/2}$ with $M_{1/2} \equiv M_1 = M_2$ at the GUT scale. 
    Then, the Bino-like neutralino is the NLSP for $M_3 \gsim 0.2 M_{1/2}$. 
    See the text for the other irrelevant parameters. 
  }
  \label{fig:TR}
\end{figure}

In Fig.~\ref{fig:TR}, we plot the maximal reheating temperature allowed by 
the overclosure bound with varying the gluino mass $M_3$ compared to the 
gaugino mass $M_{1/2}$ at the GUT scale. Here, the Bino and Wino masses 
are set to be equal $M_1 = M_2 \equiv M_{1/2}$ and realize the NLSP mass 
$m_{\rm NLSP}$ on each solid lines. Since the universal soft scalar mass 
$m_0$ except for the Higgs ones is chosen to be $m_0 = M_{1/2}$, the 
Bino-like neutralino is the NLSP, while the Higgs mass squareds are assumed 
to be $m_{\bar H}^2 = m_0^2$ and $m_H^2 = -m_0^2$ for the electroweak 
symmetry breaking to give arise especially in a large $m_0$ region. The scalar 
trilinear couplings, $a_0$, and $\tan\beta$ are $a_0 = 0$ and $\tan\beta = 30$, 
though they are irrelevant for the maximal reheating temperature. 

The gravitino mass is relevant for the upper bound on the reheating temperature. 
We assumed that it is equal to the NLSP mass. This turns out to maximize $T_R$ 
for $M_3/M_{1/2} > 0.5$. Although the possible maximal temperature is obtained 
by a smaller gravitino mass for a lighter gluino region, the difference of $T_R$ is 
less than 10\% even for $M_3/M_{1/2} = 0.2$. It is commented that the gluino 
can become the NLSP for $M_3/M_{1/2} < 0.2$. 

We find that the maximal reheating temperature can be as large as 
$O(10^8)$GeV in a large parameter region, while $T_R \geq 10^9$GeV is 
realized for a smaller NLSP mass, e.g. $m_{\rm NLSP} \lsim 600$GeV for the 
universal gaugino mass $M_3 = M_{1/2}$. This is because the thermal gravitino 
abundance (\ref{eq:gravitino-abundance}) increases as the soft mass scale is 
larger. From Fig.~\ref{fig:TR}, we see that $T_R > 10^9$GeV leads to 
$m_{\rm NLSP} < 1.5$TeV for $M_3/M_{1/2} > 0.2$ (see \cite{Hamaguchi:2009sz} 
for a hierarchical gaugino mass case). According to PAMELA, the gravitino mass is 
indicated as $m_{3/2} \gsim O(100)$GeV, {\it i.e.} $m_{\rm NLSP} \gsim O(100)$GeV. 
Then, it is easy to obtain the leptogenesis temperature in a wide class of models (see 
\cite{Buchmuller:2008vw}, in which discussions of superparticle mass spectrum are 
also explored). 

In contrast, the ATIC result implies a larger gravitino mass, $m_{3/2} \gsim 1$TeV. 
Then, the gluino mass is required to be suppressed. For instance, when the gravitino 
mass is 1TeV, $T_R > 10^9$GeV needs $M_3 \lsim 0.7 M_{1/2}$ at the GUT scale, 
and $M_3 \lsim 0.3 M_{1/2}$ for $m_{3/2} = 1.5$TeV. Thus, the SUSY breaking 
models become specified to realize the leptogenesis temperature. A lighter gluino 
mass parameter is favored at the GUT scale, and thus a mass spectrum of the 
superparticles tends to degenerate at the weak scale especially when the gravitino 
mass is heavier 
\footnote{
  This mass spectrum looks like that of the mirage mediation \cite{EYY}. 
  However, it is usually obtained by enhancing the anomaly mediation, 
  namely with a large gravitino mass, e.g. $m_{3/2} \gsim 100$TeV. 
  Thus, the gravitino is not the LSP. 
}. 
On the other hand, the universal gaugino mass models look inconsistent with the 
thermal leptogenesis in the light of the ATIC result, though the usual gauge- and 
many gravity-mediation models have this feature. Thus, we will revisit the thermal 
leptogenesis in the next section in the presence of the R-parity violation. 

Before proceeding to the next section, let us comment on the other parameter 
dependence. In the analysis, we take $m_0 = M_{1/2}$ at the GUT scale. For 
tiny $m_0$, the stau can be the NLSP. Then, the upper constraint in 
Fig.~\ref{fig:TR} becomes severer for fixed $m_{\rm NLSP}$, since the gaugino 
masses increase. Thus, the model parameters discussed above are limited 
more severely. 

%%%%%%%%%%%%%%%%%%%%%%%%%%%%%%%%%
\section{Thermal Leptogenesis Revisited}
%%%%%%%%%%%%%%%%%%%%%%%%%%%%%%%%%

We now revisit the thermal leptogenesis in the presence of the R-parity violating 
operator (\ref{eq:RHNU-RPV}). In the above, we found that a gluino mass is 
bounded from above to realize the leptogenesis temperature $T_R \gsim 10^9$GeV 
in a large gravitino mass region. In particular, the universal gaugino mass models 
are inconsistent with the temperature if we take the ATIC result. On the other hand, 
we have introduced the R-parity violating operator in the right-handed neutrino 
sector. We will see that a CP asymmetry of the decay of the right-handed neutrino 
is enhanced by the violation \cite{Farzan:2005ez}, and thus the leptogenesis 
mechanism works successfully. 

In the thermal leptogenesis, the baryon density relative to the photon density 
becomes (cf.~\cite{Buchmuller:2005eh})
\begin{equation}
  \frac{n_B}{n_{\gamma}}
  \;\simeq\; 
  -1.04 \times 10^{-2} \epsilon_1 \kappa,
\end{equation}
which is compared to the observation \cite{Hinshaw:2008kr},
\begin{equation}
  \frac{n_B}{n_{\gamma}} \;=\; (6.21\pm 0.16)\times 10^{-10}.
\end{equation}
Here and hereafter, we assume the hierarchical right-handed neutrinos and 
ignore the flavour effects which does not affect the lower bound on the reheating temperature. The efficiency factor $\kappa$ represents effects of washout and 
scattering processes, which is obtained by solving the Boltzmann equations. For 
the case of zero initial abundance of the right-handed neutrinos, its maximal value 
is known to be $\kappa \simeq 0.2$ \cite{Buchmuller:2004nz,Giudice:2003jh}. 

The CP asymmetry $\epsilon_1$ in the decay of the lightest right-handed neutrino, 
$N_1 \to LH$, is determined by the structure of the neutrino sector. With $\kappa 
\lsim 0.2$ the observed baryon asymmetry requires
\begin{equation}
  |\epsilon_1| \;\gsim\; 3 \times 10^{-7}. 
  \label{eq:eps1min}
\end{equation}
In the R-parity preserved models, it is known that the $|\epsilon_1|$ has an upper 
bound as \cite{Hamaguchi:2001gw,Davidson:2002qv,Covi:1996wh}
\begin{equation}
  |\epsilon_1|
  \;\equiv\;
  \left|
  \frac{\Gamma(N_1 \to L+H)-\Gamma(N_1 \to L^c+H^c)}
  {\Gamma(N_1 \to L+H)+\Gamma(N_1 \to L^c+H^c)}
  \right|
  \;\lesssim\;
  \frac{3M_{N1}}{8\pi\langle H\rangle^2}
  \frac{\Delta m_{\rm atm}^2}{m_1+m_3}
  \label{eq:maxeps1}
\end{equation}
due to the see-saw relation.
Here $m_i$ ($m_1 < m_2 < m_3$) are the mass eigenvalues of the light 
neutrinos, and $M_{N1}$ is the mass of the lightest right-handed neutrino $N_1$. 
Using the atmospheric neutrino mass squared difference $\Delta m_{\rm atm}^2 
\simeq (2.5 \pm 0.2) \times 10^{-3}\text{eV}^2$, the lower bound on the 
right-handed neutrino mass becomes 
\begin{equation}
  M_{N1} 
  \;\gsim\; 
  1.4\times 10^{9}\ \text{GeV} \cdot \sin^2\beta
  \label{eq:M1min} 
\end{equation}
with the 3$\sigma$ values of $n_B/n_{\gamma}$ and $\Delta m_{\rm atm}^2$. 
Consequently, the corresponding lower bound on the reheating temperature is 
obtained as $T_R \gsim 1 \times 10^9\ {\rm GeV}$. 

The upper bound on $|\epsilon_1|$ (\ref{eq:maxeps1}) can be relaxed by the 
R-parity violating operator (\ref{eq:RHNU-RPV}). The operator contributes to 
the CP asymmetric decay, $N_1 \to LH$, via the diagrams exchanging $H$ and 
$\bar{H}$ (and $N$) in the loop. In addition, a decay, $N_1 \to H\bar{H}$, gives 
arise with the R-parity violating operator of $N_1$. Consequently, we obatin 
\cite{Farzan:2005ez}
\begin{eqnarray}
  \epsilon_1^{({\rm RPV})}
  &=&
  - \frac{1}{8\pi}
  \frac{1}{(Y_NY_N^{\dagger})_{11} + |\lambda_1|^2}
  \nonumber \\
  && ~~~~~
  \times \sum_{i\neq 1}
  \mathrm{Im}
  \left[
    (Y_NY_N^{\dagger})_{i1} (\lambda_i \lambda_1^*) 
    f\left( \frac{M_{Ni}^2}{M_{N1}^2} \right)
    +
    \frac{2 (Y_NY_N^{\dagger})_{i1} (\lambda_i^* \lambda_1)} 
    {M_{Ni}^2/M_{N1}^2 - 1}
  \right]
\end{eqnarray}
where the loop function is $f(x) = -\sqrt{x}\ln(1+1/x) - 2\sqrt{x}/(x-1)$. We notice 
that the result depends on $\lambda_1$ and $\lambda_i\ (i \neq 1)$. In order to 
avoid a strong washout by a (inverse) decay, $\lambda_1$ is favored to satisfy 
$|\lambda_1|^2 \lsim (Y_NY_N^{\dagger})_{11}$, while $\lambda_i$ is allowed 
to be $O(1)$. Then, the R-parity violating processes can dominate the CP 
asymmetry of the right-handed neutrino decay. Actually, it is estimated as
\begin{eqnarray}
  \epsilon_1^{({\rm RPV})} 
  &\simeq&
  2 \times 10^{-4}\cdot 
  \mathrm{Im}[c_1^* \lambda_i]
  \nonumber \\
  &&
  \times 
  \left( \frac{M_{N1}}{10^8{\rm GeV}} \right)^{\frac{1}{2}}
  \left( \frac{M_{Ni}/M_{N1}}{10} \right)^{-\frac{1}{2}}
  \left( \frac{\tilde m_1}{10^{-3}{\rm eV}} \right)^{-\frac{1}{2}}
  \left( \frac{\bar m'_\nu}{0.1{\rm eV}} \right)
  \frac{1}{\sin^2\beta},
  \label{eq:eps1RPV}
\end{eqnarray}
for $M_{Ni}/M_{N1} \gg 1$. Here, the parameters are defined as 
$(Y_NY_N^{\dagger})_{i1} = \bar m'_\nu \sqrt{M_{N1}M_{Ni}}/v^2\sin^2\beta$, 
$\tilde m_1 = [(Y_NY_N^{\dagger})_{11} + |\lambda_1|^2] v^2 
/M_{N1}$ and $c_1 = \lambda_1 v/\sqrt{M_{N1}\tilde m_1}$. We can see that 
the result exceeds the upper bound (\ref{eq:maxeps1}) for $\lambda_i \sim 1$ 
and $|\lambda_1|^2 \lsim (Y_NY_N^{\dagger})_{11}$, i.e. $c_1 \sim 1$. 

In the last section, we found that $T_R = O(10^8)$GeV is obtained even in a 
heavy NLSP mass region from Fig.~\ref{fig:TR}. On the other hand, we saw 
in this section that the CP asymmetry of the right-handed neutrino can be 
drastically enhanced by the R-parity violating operator (\ref{eq:RHNU-RPV}), 
and the condition (\ref{eq:eps1min}) is easily satisfied, e.g. for $M_1 = 10^8$GeV. 
Since $N_1$ is expected to be produced sufficiently in the thermal bath once 
$T_R$ exceeds $M_{N1}$, the thermal leptogenesis consequently works in a 
wide class of models with explaining the cosmic-ray anomalies from PAMELA/ATIC. 
To be explicit, let us give an example: $\lambda_{i \neq 1} \sim 1$ with the lightest 
right-handed neutrino mass $M_{N1} = O(10^8)$GeV and $M_{Ni}/M_{N1} = O(10)$. 

One may be worried that such a large R-parity violation enhances washout effects.
The (inverse) decay and $\Delta L = 1$ processes indeed receive additional 
contributions from the R-parity violating operator. Considering that they are 
controlled by the washout mass parameter, $\tilde m_1 = (Y_NY_N^{\dagger})_{11}
v^2/M_{N1}$, we notice that the extra effects just redefine $\tilde m_1$ as the new 
one given in the previous paragraph. On the other hand, the R-parity violation can 
enhance the $\Delta L = 2$ washout. Actually, we obtain $LH \to HH$ ($H$  means 
$H$ or $\bar{H}$) by exchanging a heavier right-handed neutrino in the diagram. 
Since the diagram is almost the same as the R-parity preserved one except for a 
coupling and corresponding field, the additional contribution is roughly estimated as 
\begin{eqnarray}
  \gamma_{\rm RPV} 
  \;\sim\;
  \gamma_{\Delta L = 2}^{(sub)}
  \times
  \frac{ |\lambda_i|^2 |(Y_N)_{i1}|^2 }{ |(Y_N)_{11}|^4 }
  \frac{M_{N1}^2}{M_{Ni}^2},
\end{eqnarray}
where $\gamma_{\Delta L = 2}^{(sub)}$ represents the $\Delta L = 2$ washout 
term with a subtraction of the resonance \cite{Giudice:2003jh}. Thus, 
$\gamma_{\rm RPV}$ becomes larger than $\gamma_{\Delta L = 2}^{(sub)}$ for 
$M_{N1} \ll 10^{13}$GeV. Since $\gamma_{\Delta L = 2}^{(sub)}$ is negligibly small 
for this $M_{N1}$, we can numerically check that the washout is dominated by the 
(inverse) decay and $\Delta L = 1$ scattering even for $\lambda_i = O(1)$ (see 
e.g. \cite{Giudice:2003jh} for the numerical estimations of the washout terms in the 
R-parity preserved case). Although $\gamma_{\Delta L = 2}^{(sub)}$ increase as 
$M_{N1}$ grows, $\gamma_{\rm RPV}$ decreases at the same time. Thus, the 
R-parity violating contribution does not change the result.

%%%%%%%%%%%%%%%%%%%%%%%%%%%%%%%%%
\section{Conclusions}
%%%%%%%%%%%%%%%%%%%%%%%%%%%%%%%%%

The gravitino DM scenario is an attractive candidate to explain the PAMELA/ATIC 
excesses. However, the R-parity violation has to be tightly suppressed in the 
low-energy scale. In this letter, we proposed a framework of the R-parity violation 
to solve the fine-tuning. Introducing the violation in the right-handed neutrino 
sector, the effects are naturally suppressed in the low-energy scale by the see-saw 
mechanism. We found that the life time required by PAMELA/ATIC, $\tau_{3/2} = 
O(10^{26})$sec, is obtained by the R-parity violating operator $W = \lambda 
NH\bar{H}$ with $\lambda \sim 1$ for $M_N = O(10^9)$GeV. 

The R-parity violation not only provides a well source of the cosmic rays to explain 
the PAMELA/ATIC anomalies, but also opens a window for the thermal leptogenesis. 
In the presence of the R-parity violation, the reheating temperature is bounded from 
above by the overclosure bound due to the thermal gravitino production. We found 
that $T_R \gsim 10^9$GeV is accessible for $m_{3/2} < 600$GeV, while $m_{3/2} 
\gsim 1$TeV favored by ATIC restricts the gluino mass to be suppressed for $T_R 
\gsim 10^9$GeV. Nonetheless, the leptogenesis was shown to work, since the 
R-parity violating interactions in the right-handed neutrino sector enhances the CP 
asymmetric right-handed decay. Therefore, the gravitino LSP with the R-parity 
violation in the right-handed neutrino sector is a nice candidate of the decaying DM 
scenarios. 

It is worthwhile to clarify the parameter dependence of the phenomena. 
The cosmic-ray spectra are sensitive to larger $\lambda_2$ and corresponding 
$M_{N2}$ but insensitive to $M_{N1}$ as long as $\lambda_1 \ll \lambda_2$, 
while the lightest right-handed neutrino determines the thermal leptogenesis, 
{\it i.e.} the baryon asymmetry depends on $M_{N1}$. From (\ref{eq:eps1RPV}) 
it is possible to lower $M_{N1}$ as well as the reheating temperature required 
by the thermal leptogenesis, though too small $M_{N1}$ needs tiny $\lambda_1$ 
to avoid a strong washout. 

The cosmic-ray spectra of the anti-matter and photon also depends on the 
R-parity violation. In this letter, the flavor structure is determined by the neutrino 
Yukawa coupling and $\lambda^i$. By obtaining more cosmic-ray data in future 
such as FGST and AMS-02 and refining the astrophysical knowledge, we may 
distinguish the R-parity violating operators. Furthermore, since the R-parity 
violating operator can be embedded in a high-scale models such as SU(5) GUT 
and heterotic string models, combining with the future collider data we might be 
able to study physics in high-energy scale.

%%%%%%%%%%%%%%%%%%%%%%%%%%%%%%
\section*{Acknowledgment}
%%%%%%%%%%%%%%%%%%%%%%%%%%%%%%
The authors thank M.~Kakizaki  for discussions on the R-parity violation, 
D.~Tran for useful comments on the cosmic-ray, and W.~Buchm\"uller 
for reading the manuscript. 

%%%%%%%%%%%%%%%%%%%%%%%%%%%%%%%%%

%%%%%%%%%%%%%%%%%%%%%%%%%%%%%%%%%

\end{document}